\documentclass[prd,preprint]{revtex4}
\usepackage{graphicx}
\usepackage{amsmath,amssymb}
\usepackage{epsfig}
\usepackage{latexsym}
\newcommand {\ga} {\ {\raise-.5ex\hbox{$\buildrel>\over\sim$}}\ }
\newcommand {\la} {\ {\raise-.5ex\hbox{$\buildrel<\over\sim$}}\ }

\begin{document}

\def\be{\begin{equation}}
\def\ee{\end{equation}}

\title{Implications of Broken Symmetry for Superhorizon Conservation Theorems in Cosmology}
\author{Katherine Jones-Smith$^1$, 
Lawrence M. Krauss$^{2}$ and Harsh Mathur$^1$ }
\affiliation{$^1$CERCA, Department of Physics, Case Western Reserve University,
Cleveland, OH~~44106}
\affiliation{$^2$School of Earth and Space Exploration, Physics Department, and Beyond Center, Arizona State University, Tempe, AZ 85287}
\date{\today}

\begin{abstract}

Inflation produces super-horizon sized perturbations that ultimately return within the horizon and are thought to form the seeds of all observed large scale structure in the Universe.  But inflationary predictions can only be compared with present day observations if, as conventional wisdom dictates, they remain unpolluted by subsequent sub-horizon causal physical processes  and therefore remain immune from the vicissitudes of unknown universal dynamics in the intervening period.   Here we demonstrate that conventional wisdom need not be correct, and as a result cosmological signatures arising from intervening unknown non-inflationary processes may confuse the interpretation of observational data today.

\end{abstract}

\maketitle

The prospect that astrophysical observations may allow us to directly detect the signatures of processes relevant at the earliest moments of the big bang continues to be one of the most exciting possibilities in cosmology.  The key question remains: to what extent are primordial signatures polluted by intervening processes during the subsequent evolution of the universe?  

One might hope that super-horizon sized signatures are preserved until these scales enter the horizon and causal processes may then affect them. 
Beginning with the seminal work of Bardeen in 1970's the analysis of super-horizon sized modes in cosmology has focused on conservation laws that might permit one to extrapolate back from 
present-day observations to directly connect with the physics of the primordial universe
\cite{bardeen, lyth, steinhardt, liddle}. 
With the advent of inflationary theories, which actually provide calculable initial conditions from which one might hope to understand the formation of large scale structure, and which also involve initial sub-horizon modes becoming super-horizon sized, the motivation for this program increased tremendously. Can one extrapolate forward by a tremendous amount these initial conditions through a period in which the contents of the universe and the local dynamics are unknown in order to make contact with observations made during the present epoch?   

The common presumption that this is indeed the case culminated in a series of theorems by Weinberg, who proved the existence of solutions for which appropriately defined amplitudes of superhorizon tensor and scalar modes remain constant until they re-enter the horizon, independent of the contents of the universe and the possibly unknown laws that govern dynamics during the period between exit and re-entry of the horizon \cite{weinberg1, weinberg2, weinbergtext,weinberg3}.

Nonetheless the key issue remains whether the solutions that are guaranteed to exist by Weinberg's theorem are always the relevant ones. (A complementary, but distinct question, which has been studied in the literature, is whether the overall spatial pattern of CMB temperature anisotropies predicted by
inflation is unique to inflation, or whether it can be reproduced, at least in
principle, by local causal processes \cite{turok1, turok2, hu}.)

In this Letter we show that while the proof of Weinberg's theorem is mathematically correct, there is at least one important physical circumstance that obviates it, namely symmetry breaking. Under this circumstance modes can undergo substantial evolution even outside the horizon, with significant physical consequences that may require a reinterpretation of cosmological observations.

In the analysis of cosmological perturbations it is assumed that to a first approximation
the Universe is homogeneous and isotropic and its expansion described by a single
scale factor $a(t)$. The corrections to this geometry are described by the metric perturbation
$h_{\mu \nu}$, called the strain tensor, and customarily decomposed into scalar, vector
and tensor components. The particles and fields that constitute the Universe are likewise
assumed to be homogeneously distributed to first approximation. Their deviation from 
homogeneity is described by the stress tensor $\Pi_{\mu \nu}$, which may likewise be
decomposed into scalar, vector and tensor components. The evolution of the
strain is driven by the stress in accordance with the linearized Einstein equation. 
The evolution of the stress on the other hand is determined by the laws that govern
the particles and fields that constitute the Universe. It is this information that is currently lacking
for a significant duration between the end of inflation and the appearance of phenomena
that are now observable such as the cosmic microwave background. 

At first sight it might appear impossible to say anything about the evolution of the
Universe in the absence of information about the laws the govern its constituents. 
However these unknown laws are in fact constrained by general covariance or
equivalently the gauge invariance of linearized gravity. Using this invariance, Weinberg proved
that no matter what laws govern the constituents of the 
Universe, there always exist solutions for which appropriately defined spatial Fourier
amplitudes of the strain at super-horizon wave-vectors remain constant in time.
Whether these solutions, which are mathematically allowed, are actually realized of course 
depends on the initial conditions that apply at the end of the inflation. Since these solutions
agree with the intuition that modes do not evolve outside the horizon the conventional wisdom
has been that they are physically relevant, but we will now argue that presumption is not 
necessarily justified.

We explore here a specific example for which quantitative results are available \cite{jones-smith}, re-examined in light of Weinberg's theorem. 
Consider a scalar field that undergoes a continuous
symmetry breaking phase transition somewhat after inflation.   
For simplicity we consider only the tensor sector though similar considerations should
apply in the scalar sector also.  We assume that shortly after inflation the field becomes thermally
disordered.  The field then orders as it relaxes; by virtue of causality it can only be ordered on horizon scales.  It is important to note in this regard that this disorder on super-horizon sizes means the field cannot be spatially decomposed into a constant term plus small perturbations (i.e see  \cite{weinbergbook2}), which in turn results in a time varying anisotropic stress tensor which can be a source of
gravitational radiation (corresponding to tensor perturbations in the gravitational field) \cite{krauss}. 
Thus at any given time there is stress on the horizon scale
and therefore the strain spectra are peaked around the 
horizon wave-vector. But then by continuity there is some also weight at larger, super-horizon 
modes. Thus if we focus on a particular super-horizon wave-vector it is
clear that the strain amplitude at that wave-vector will grow in time, reaching a peak
when the mode comes into the horizon. Therefore, for the physically
reasonable solution in the event of symmetry breaking, the super-horizon tensor 
modes grow with time. 

As for the guaranteed existence of solutions where the gravitational modes do not
grow, such solutions still do exist when the constituents of the universe include
symmetry breaking fields; however these solutions can be somewhat artificial. 
For example, if thermal or quantum mechanical disordering after inflation is turned off so that the
scalar field is initially perfectly ordered, it will then remain
ordered for all times.  In this case it will not produce gravitational radiation on any scale, super- or
sub-horizon.

The general physical arguments given above are borne out by explicit calculations. 
We focus on the
tensor perturbation $D_{ij}$ that corresponds physically to gravitational radiation and
is related to the strain via $h_{ij} = a^2 D_{ij}$. The perturbation is symmetric
($D_{ij} = D_{ji}$), traceless ($D_{ii} = 0$) and transverse ($ \partial_i D_{ij} = 0$). 
It is convenient to expand the spatial Fourier amplitude
$D_{ij} ({\mathbf q}, \tau) = \sum_{\lambda= \pm 2} D_{\lambda} ({\mathbf q}, \tau) e_{ij} (\hat{\mathbf q}, \lambda)$
where $D_{\lambda} ({\mathbf q}, \tau)$ is the amplitude of a gravitational wave of co-moving wave-vector
${\mathbf q}$ and polarization $\lambda$ at conformal time $\tau$. 
The matrix $e_{ij} (\hat{\mathbf q}, \lambda)$ for
propagation along an arbitrary direction may be obtained by rotating the tensor for propagation 
along the $z$-axis; the non-vanishing elements of the latter are 
$e_{xx} ({\mathbf z}, \pm 2) = e_{yy} ({\mathbf z}, \pm 2) = 1/\sqrt{2}$ and 
$e_{xy} ({\mathbf z}, \pm 2) = e_{yx} ({\mathbf z}, \pm 2) = \pm i/\sqrt{2}$.
The evolution of the amplitudes $D_{\lambda} ({\mathbf q}, \tau)$ is governed by the linearized
Einstein equation,
\begin{equation}
\frac{\partial^2}{\partial \tau^2} D_{\lambda} + \frac{2}{a} \frac{\partial a}{\partial \tau} 
\frac{\partial}{\partial \tau} D_{\lambda} 
+ q^2 D_{\lambda}
= 16 \pi G \Pi_{\lambda}^{T},
\label{eq:einstein}
\end{equation}
essentially the wave equation in an expanding Universe. Here the source term
$\Pi_{\lambda}^T$ is the traceless transverse component of the stress perturbation.
Weinberg's theorem asserts that no matter what laws govern the Universe there
exists a solution such that the tensor amplitudes $D_{\lambda} ({\bf q}, \tau)$ remain
constant for super-horizon wave-vectors $ q/a \ll \dot{a}/a$. 
In effect, the theorem says that there always exists a solution in which the super-horizon stress 
vanishes, {\em i.e.} $\Pi^T_{\lambda} ({\bf q},\tau)  \approx 0$ for $q /a \ll \dot{a}/a$.

To determine the strain that actually results when there is symmetry breaking we consider
a particular model, the $O(N)$ non-linear sigma model wherein 
$N$ massless free scalar fields $\phi_{\alpha}$ are coupled by the constraint
$\sum_{\alpha=1}^{N} \phi_{\alpha}^2 = \eta^2$. 
Here $\eta$ denotes the vacuum expectation value of the ordered
field. We assume that the field is initially disordered and therefore impose white noise
initial conditions. The subsequent evolution of the field was worked out analytically
in the large $N$ limit in reference \cite{jones-smith}; the results are simplest when the scale factor
varies as a simple power law  of conformal time $a = (\tau/\tau_n)^{\beta}$ where
$\tau_n$ is the present conformal time. The key finding is that the two point correlator
$\langle \phi_{\alpha} ({\mathbf k}, \tau_1) \phi_{\gamma} ({\mathbf p}, \tau_2) \rangle
= (2 \pi)^3 \delta_{\alpha \gamma} \delta ({\mathbf k} + {\mathbf p}) C( k, \tau_1, \tau_2 )$
has a scaling form and moreover four-point and higher correlators may be expressed in
terms of the two-point correlator via Wick's theorem. 

For example the equal time two-point correlator has the form
$C(k, \tau, \tau) = \tau^3 f(k \tau)$ with the scaling function $f(x) \rightarrow 0$ for 
$ x \gg 1$ and approaching a constant value for $x \ll 1$; this form can easily
be understood to reflect the ordering of the scalar field on sub-horizon scales.  
It follows from the scaling form and simple power counting that the typical stress
$\Pi^T_{\lambda} ({\mathbf q}, \tau) \sim 1/\sqrt{\tau}$ for short times $\tau \ll q$. 
Moreover in the short time 
limit we may neglect the damping and restoring terms (the second and third terms on the 
left hand side) of the linearized Einstein eq (\ref{eq:einstein}), leading to the conclusion
$D_{\lambda} ({\mathbf q}, \tau) \sim \tau^{3/2}$ for short times $\tau \ll q$ confirming 
that the ordering of the symmetry breaking field causes the tensor modes to grow
outside the horizon. The full calculation yields the result that
the power in the tensor modes outside the horizon ($\tau \ll q$) grows as
\begin{equation}
P( q, \tau) = C_{\beta} G^2 \eta^4 ( p \tau )^3/N.
\label{eq:power}
\end{equation}
Here the power is defined by
$2 \pi^2 P(q, \tau) \delta( {\mathbf q} + {\mathbf p} ) = 
q^3 \sum_{\lambda= \pm 2} \langle D_{\lambda} ({\mathbf q}, \tau) D_{\lambda} ({\mathbf p}, \tau)
\rangle$. $G$ is Newton's constant, and the constant $C_{\beta} = 2500$ for $\beta = 1$ (appropriate for a radiation
dominated Universe). 

Although we have not considered the case in which the ordering dynamics
of the symmetry breaking field is frustrated by the formation of topological defects such 
as monopoles and strings, it is clear on physical grounds that in these cases too symmetry
breaking will drive growth of the super-horizon modes. Also, although we have confined
our analysis to the tensor sector for simplicity, it is expected that similar considerations 
should apply to the scalar sector. 

Thus, while Weinberg's theorem remains formally correct, in the case of symmetry breaking it does not constrain the growth of every possibly physically significant mode, and thus it does not guarantee the direct observability of inflationary signatures in all cases.  This issue may become important in the interpretation of upcoming CMB observations.

We acknowledge valuable discussions and correspondence with Steven Weinberg.


\begin{thebibliography}{99}
\bibitem{bardeen} J.M. Bardeen, Phys Rev {\bf D22}, 1882 (1980).
\bibitem{lyth} D.H. Lyth, Phys Rev {\bf D31}, 1792 (1985).
\bibitem{steinhardt} J.M. Bardeen, P.J. Steinhardt, and M.S. Turner, Phys Rev {\bf D28}, 679 (1983).
\bibitem{liddle} D. Wands, K.A. Malik, D.H. Lyth, and A.R. Liddle, Phys Rev {\bf D62}, 043527 (2000).
\bibitem{weinberg1} S. Weinberg, Phys Rev {\bf D67}, 123504 (2003); \newline arXiv/0302326.
\bibitem{weinberg2} S. Weinberg, Phys Rev {\bf D69}, 023503 (2004); \newline arXiv/0306304.
\bibitem{weinbergtext} S. Weinberg, {\em Cosmology} (Oxford University Press, 2008), Sec. 5.4.
\bibitem{weinberg3} S. Weinberg, arXiv:0808.2909; arXiv:0810.2831.
\bibitem{turok1} N. Turok, Phys. Rev. {\bf D54}, R3686 (1996).
\bibitem{turok2} N. Turok, Phys  Rev. Lett. {\bf 77}, 4138 (1996).
\bibitem{hu} W. Hu, D.N. Spergel and M. White, Phys Rev {\bf D55}, 3288 (1997).
\bibitem{jones-smith} K. Jones-Smith, L.M. Krauss and H. Mathur, Phys. Rev. Lett. {\bf 100}, 131302
(2008).
\bibitem{weinbergbook2} S. Weinberg, {\em Cosmology} (Oxford University Press, 2008), p. 499
\bibitem{krauss} L. M. Krauss, Phys. Lett. {\bf B284}, 229 (1992)
\end{thebibliography}
\end{document}